# A multilayer interstitial fluid flow along vascular adventitia

**Hongyi Li** [1, 2 *], **You Lv** [2], **Xiaoliang Chen** [4], **Bei Li** [2], **Qi Hua** [1 *], **Fusui Ji** [2 *], **Yajun Yin** [5], **Hua Li** [6]

[1] Cardiology Department, Xuanwu Hospital, Capital Medical University, No. 45, Changchun Street, Xicheng, Beijing, China

[2] Cardiology Department, [3] Radiology Department, Beijing Hospital, National Center of Gerontology, Institute of Geriatric Medicine, Chinese Academy of Medical Science, Beijing, China

[4] Radiology Department, China-Japan Friendship Hospital, Beijing Hospital, China

[5] Department of Engineering Mechanics, Tsinghua University, Beijing, China

[6] Institute of Computing Technology, Chinese Academy of Sciences, Beijing, China

**Corresponding author:**

Hongyi Li and Fusui Ji, Cardiology Department, Research Center for Interstitial Fluid Circulation & Degenerative Diseases and Aging, Beijing Hospital, Beijing 100730, China. Email: leehongyi@bjhmoh.cn (HyL) and jifusui0367@bjhmoh.cn (FsJ)

Qi Hua, Cardiology Department, Xuanwu Hospital, Capital Medical University, No. 45, Changchun Street, Xicheng, Beijing 100053, China. Email: huaqi5371@medmail.com.cn

**Abstract:**

**Objective**: Interstitial fluid flow through vascular adventitia has been disclosed recently. However, its kinetic pattern was unclear.

**Methods and Results**: We used histological and topographical identifications to observe ISF flow along venous vessels in rabbits. By MRI in alive subjects, the inherent ISF flow pathways in legs, abdomen and thorax were enhanced by paramagnetic contrast from ankle dermis. By fluorescence stereomicroscopy and layer-by-layer dissection after the rabbits were sacrificed, the perivascular and adventitial connective tissues (PACT) along the saphenous veins and inferior vena cava were found to be stained by sodium fluorescein from ankle dermis, which coincided with the findings by MRI. By confocal microscopy and histological analysis, the stained PACT pathways were verified to be the fibrous connective tissues and consisted of longitudinally assembled fibers. By usages of nanoparticles and surfactants, a PACT pathway was found to be accessible for a nanoparticle under 100nm and contain two parts: a tunica channel part and an absorptive part. In real-time observations, the calculated velocity of a continuous ISF flow along fibers of a PACT pathway was 3.6-15.6 mm/sec.

**Conclusion**: These data further revealed more kinetic features of a continuous ISF flow along vascular vessel. A multiscale, multilayer, and multiform "interstitial/interfacial fluid flow" throughout perivascular and adventitial connective tissues was suggested as one of kinetic and dynamic mechanisms for ISF flow, which might be another principal fluid dynamic pattern besides convective/vascular and diffusive transport in biological system.

# Introduction

Vascular vessels consist of three layers: tunica intima, tunica media and tunica adventitia. They produce an intraluminal space to rapidly transport $O_2$, nutrients, waste products and heat around the body. The outermost tunica adventitia, together with the surrounding perivascular/paravascular loose connective tissues, serves mainly to strengthen blood vessel and anchor it to nearby organs, giving stability. However, increasing data demonstrated that both the perivascular and adventitial connective tissues (PACT) are capable to transport interstitial fluid (ISF) as well [1-3].

ISF flow along venous adventitia was found along the lower extremity veins, inferior vena cava of abdomen and thorax, into three grooves of the heart and pericardial cavity in rabbits [1]. Both tunica adventitia and its surrounding fibrous connective tissues participated in ISF flow along blood vessel walls [1]. In humans, ISF flow along blood vessels were identified to be along not only venous adventitia but also arterial adventitia [2, 3]. Histologically, vascular adventitia and its surrounding tissues are the fibrous connective tissues [1-3]. We named it as adventitial ISF flow that included at least two types: fluid flow through tunica adventitia and the perivascular/paravascular connective tissues.

Another pattern of fluid flow along blood vessels is a perivascular space (PVS) that has been noticed in brain for centuries. The PVS is distributed between the outer and inner lamina of brain vessels or around fenestrated capillaries. Recently, a brain-wide perivascular pathway was identified and named as Glymphatic system, which includes the periarterial influx of cerebrospinal fluid (CSF) into the brain interstitium, followed by the clearance of ISF along large-caliber draining veins, serving as a lymphatic-like clearing system in the brain [4, 5]. However, a Glymphatic pathway is in the vicinity of tunica adventitia of vascular conduits [4]. The relationship of Glymphatic and ISF flow through a PACT pathway has not been disclosed yet.

In conventional interstitial physiology, the matrix of fibrous connective tissues is a fluid-filled porous/interstitial media with a random net of internal pores/interstices. Based on the random porous matrix, adventitial ISF would flow diffusively under a pressure/concentration gradient other than a long-distance transport [6]. However, our previous data revealed that the solid structure of fibrous matrices is a highly ordered fibrous framework that contains abundant oriented fibers [3]. Because neither the fiber nor the gel themselves can flow, the only possible transport space for ISF might be an *interfacial zone* (or interfacial transport zone, **ITZ**) between a solid phase (a fiber) and a liquid phase (the gel/liquid substance) [2, 3]. When the fluorescent ISF enters an ITZ along a fiber, the oriented fibers of fibrous matrices would work as a fibrorail and be fluorescently stained [7, 8]. Thus, a long-distance ISF flow pathway can be identified by tracking the fluorescently stained fibers via Fluorescence or Confocal microscopy.

In order to explore the kinetic features of the long-distance ISF flow along blood vessels in the body, we designed the following experiments to observe the movements of imaging tracers to represent ISF flow through a PACT pathway in rabbits, which were the paramagnetic contrast, sodium fluorescein, a solution mixed with different surfactants and nanoparticles respectively.

**Materials and Methods**

*Subjects*

The study was conducted in Beijing between January 2011 and August 2019. A total of 67 New Zealand White rabbits (body weight 2-3 kg) were involved, and the study protocol was approved by the animal ethics committee of the Institute of People's Hospital of Peking University (No. 201170). Anesthesia (pentobarbital at 25-30 mg/kg/hour) was administered intravenously before each of the following experiments. The tracer solution with 2% lidocaine was used for local anesthesia. Each subject was sacrificed via an overdose of pentobarbital.

*Methods of visualizing the PACT pathways by MRI*

The $1^{st}$ group of 5 rabbits were examined by MRI scanning. Gd-DTPA (diluted to 156 mg/mL by physiological sodium saline) and fluorescein sodium solution (FluoNa, diluted to 10 mg/dL by physiological sodium saline) were mixed. 0.1-0.2 mL of mixed tracers was injected hypodermically into ankle dermis by a 1-ml syringe (B. Braun Medical, Shanghai). Beneath ankle dermis, there were few arterial and venous vessels passing through. The imaging tracers were injected into the hypodermic tissues but not the intravascular cavity and represented an upstream site of the long-distance PACT pathway along blood vessels.

The imaging was performed in an Achieva 3.0T TX scanner (Philips Electronics, China). Each of the subjects was in real-time scanned approximately 45–60 minutes after the hypodermic injection of Gd-DTPA. The images were obtained using a 3D T1-weighted fast field echo (FFE) sequence with an 8-channel phased-array head surface coil. Scanning parameters were adjusted to obtain a higher spatial resolution. The acquisition time for each sequence was approximately 4-6 minutes. The raw data were analyzed at the extended MR WorkSpace station with multiplanar reconstruction (MPR) and maximum-intensity projection reconstructions (MIPs). After the MRI scans, the rabbits were sacrificed and examined by fluorescence stereomicroscopy. The fluorescent images were compared with the reconstructed images by MIPs.

*Methods of visualizing the PACT pathways by fluorescence stereomicroscopy*

The $2^{nd}$ group of 5 rabbits were examined by fluorescence stereomicroscopy. Firstly, incisions were made to form a flap revealing the great saphenous vein (GSV) or the small saphenous vein (SSV) under the facial plane on the level of the thighs. Secondly, 0.1-0.2 mL of fluorescein sodium water solution (diluted to 10 mg/dL by physiological sodium saline) was injected hypodermically into the same area of the ankle dermis in 1$^{st}$ group. Photos of the fluorescent ISF via the PACT pathways of GSV or SSV were recorded in real-time by stereomicroscopy with CCD or a digital camera after the administration of FluoNa.

To elucidate the differences between ISF flow in the PACT pathways and blood flow in the intraluminal cavity of blood vessel, extra surgical incisions were performed in 5 rabbits in the 3$^{rd}$ group. Firstly, an arch-shaped incision was made in a PACT pathway of SSV, including layers of covering fascia and parts of tunica adventitia of the vessel by ophthalmic scissor. Secondly, a small branch of a main vein in ankle was punctured by 1-ml syringe and 20-30 μl FluoNa water solution was added into blood. Photos of SSV were recorded in real-time by fluorescence stereomicroscopy for 3 min observation. Thirdly, 0.1-0.2 mL FluoNa water solution was dropped on the exposed main vein. The results were recorded for the following 60-90 min.

*Observations of the PACT pathways by Confocal Laser Scanning Microscopy (CLSM) or Scanning Electron Microscopy (SEM)*

4 rabbits of 4$^{th}$ group were administrated by fluorescence and studied by CLSM or SEM. 1 rabbit of 4$^{th}$ group was administrated by physiological saline as control. All 5 subjects were sacrificed at 15 minutes after the hypodermic injection. The fluorescently stained tissues were sampled, including the segments of the SSV, GSV, inferior vena cava (IVC), anterior interventricular groove, posterior interventricular groove, coronary groove on the heart, and the right upper pulmonary veins. The same samples sampled in the control rabbit. All samples in 2$^{nd}$, 3$^{rd}$, 4$^{th}$ group were examined by CLSM, SEM or TEM.

*Examination of the PACT pathways by histological methods*

The cross-sections of the fluorescently stained PACT pathways in the 2$^{nd}$, 3$^{rd}$, 4$^{th}$ group were studied by hematoxylin & eosin (HE) staining and the combined staining methods of Van Gieson & Verhoeff Iron-Hematoxylin (VG+VIH).

*The power of "push" and "pull" in driving fluid transport through the PACT pathway in vivo*

To illustrate the driving force of fluid through the PACT pathways, an isolated segment of the PACT pathway on SSV or GSV was studied in the 5 rabbits of 5$^{th}$ group. Before the following experiments, both the distal and proximal sides of a segment of GSV or SSV were ligated by sutures to isolate the segmental PACT pathway between the distal and proximal side of the vein from the surrounding tissues.

First, both the distal end (upstream) and proximal end (downstream) of a venous vessel were ligated and the blood stopped completely (Figs. 5A1, 5A2, 5B1, 5B2). Then, 5 μl of FluoNa water solution was dropped on the middle of the ligated segment to observe the movement of FluoNa along the PACT pathway on the vein. Second, while untying the proximal ligation of the ligated vessel and keeping the distal end ligated (Fig. 5B3), the movement of FluoNa along the PACT pathway was recorded. Third in a different vessel, while untying the distal ligation of the ligated vessel and keeping the proximal end ligated (Fig. 5A3), the movement of FluoNa along the PACT pathway was recorded. During these experiments, the isolated vessel wall was kept moist by adding drops of physiological saline. Fluid flow through the segmental PACTs was recorded by fluorescence stereomicroscopy.

*Fluid transport through the PACT pathways in the dead rabbits by the mechanical manipulations on the heart*

The 6th group of 5 rabbits were sacrificed by an injection of 20 mL of air via the auricular veins. Six hours after the sacrifice, all 5 rabbits were prepared as follows. Firstly, a 10-cm-long segment of the left SSV was exposed to air in all 5 subjects. Secondly, 2 rabbits were selected and 50 μl of collagenase type I solution (diluted to 10% by physiological saline) were administered in the proximal end of the left SSV. Thirdly, the thoracic cavities of all 5 sacrificed rabbits were opened and the heart apexes were cut open to ensure the right and left ventricular chambers were exposed to air.

After the above preparations, 0.1ml of FluoNa water solution was injected hypodermically into the ankle dermis of the exposed left SSVs. Without extra manipulations, the movement of FluoNa along the PACT pathways was observed for 60 min. After the 60 min observation, the exposed heart was compressed repeatedly according to the standard procedure of open-chest cardiac compression at the frequency of 60-80 bpm for 20 min. The movements of the FluoNa were continuously photographed by stereomicroscopy. The fluorescently stained PACTs sampled from the left SSV, the IVC in the splanchnocoele, the coronary groove, and the anterior and posterior ventricular grooves were also studied by LCSM.

*Examination of the transport of nanoparticles with various sizes through the PACT pathways in alive rabbits*

We used two types of nanoparticles with different sizes to observe whether the nanoparticles could be transport via the PACT pathways in *in vivo* rabbits, which were the gold nanoparticles in 5 rabbits of 7th group and fluorescent polystyrene (PS) spheres in the 5 rabbits of 8th group. A 10-cm-long segment of the PACT pathway of both left and right SSV was exposed in air in each rabbit of the 7th, 8th group.

In 7th group, 10 μl PBS solution mixed with the five types of colloidal gold nanoparticles (diameters of 5 nm, 10 nm, 20 nm, 40 nm, and 60 nm) was dropped into the distal end of the PACT pathways in each subject. The proximal end of SSV were sampled at 10 minutes after the administration and subsequently examined by Transmission Electron Microscope (TEM).

In 8th group, 10 μl PBS solution mixed with two types of fluorescent PS spheres (diameters of 30nm and 100nm) was dropped into the distal end of the PACT pathways in each subject. The movements of fluorescent PS spheres through the exposed SSV were recorded by fluorescence stereomicroscopy.

*Observing the movements of a mixed solution of FluoNa and surfactant through the PACT pathways*

The 9th group of 5 rabbits were studied to observe the movements of a mixed solution of FluoNa and a surfactant. The mixed solution was a water solution of FluoNa (diluted to 156 mg/mL by physiological saline) and an anionic surfactant, sodium dodecyl sulfate (SDS) (diluted to 104.00 mmol/L by physiological saline at 25°C). A 10-cm-long segment of the PACT pathway of both the SSV and GSV was exposed to air in each rabbit of the 9th group.

0.1-0.2 mL mixed tracers were injected hypodermically into the ankle dermis in each subject. The movements of this mixed tracers through the PACT pathways of the exposed GSV or the SSV were recorded in real-time observation by fluorescence stereomicroscopy in all 5 rabbits.

*Observing the movements of another nonionic and cationic surfactants through the PACT pathways*

In order to observe the effects of different types of surfactants on the movements of FluoNa through the PACT pathways, the other 10 rabbits were divided equally into the 10th, 11th group. The surfactants were the nonionic surfactant, Trixon-100 and cationic surfactant, hexadecyltrimethylammonium bromide (CTAB) respectively. Each type of surfactants was tested in 5 rabbits of each group. The concentrations at 25°C were 27.60 mmol/L Trixon-100 and 7.20 mmol/L CTAB (diluted by physiological sodium solution at 25°C). The concentration of each solution was 6 times of the critical micelle concentration of each surfactant at 25°C. The experimental procedures were the same as those for SDS in 9[th] group. The movements of each fluorescent surfactant solution through the PACT pathways of the exposed GSV or the SSV were photographed in real-time observation by fluorescence stereomicroscopy.

*Comparison of the pericardial fluid between the physiological saline solution and the SDS solution by echocardiography*

A total of 3-4 mL FluoNa water solution in 3 rabbits of 12[th] group and 3-4 mL FluoNa SDS solution in the other 3 rabbits of 13[th] group were injected hypodermically into the left and right ankle dermis of each group as we introduced previously [1]. Immediately afterwards, the pericardial fluid of each subject was in real-time detected by echocardiography. After the echocardiography, the rabbits were sacrificed, and their thoracic cavities were opened to examine the amount of pericardial fluid.

*Observing the movements of ISF through the PACT pathways under higher concentration of SDS solution*

To observe whether ISF transport through the PACT pathways could be interrupted by the SDS solution with higher concentration, 6 rabbits in the 14[th] group were studied. 5%, 10%, 20% SDS solution (dilated in physiological saline) were prepared and respectively used in every 2 subjects. The GSV of each subject was exposed by surgery and bathed in SDS solution while the rabbits took supine position. 0.1-0.2 mL FluoNa water solution was injected hypodermically into ankle dermis. The FluoNa transport processes via the PACT pathway along the GSV in different SDS solution was recorded by real-time fluorescence microscopy.

## Discussion and Conclusions

The kinetics of ISF flow along vascular vessels was not clear yet. Due to the lack of a direct imaging technique and the complexities of multiscale measurements for fluid flow in fibrous connective tissues, we used multiscale and topographical approaches to explore ISF flow through a venous PACT pathway in rabbits.

*The inherent PACT pathways for ISF flow in the body that were visualized by MRI and fluorescence*

*stereomicroscopy*

By the paramagnetic Gd-DTPA in the 1st group, the panorama of the enhanced long-distance PACT pathways were clearly visualized at macroscopic level, which were located in the lower limbs, along the saphenous veins, the IVC of the abdominal and thoracic cavities and into some tissues of the heart (Figs. 1A1, 1A2, Video 1). However, the detailed anatomical and histological structures of the PACT pathways cannot be revealed by the MRI technique. Tracing the FluoNa stained tissues under a fluorescence stereomicroscopy in the 2nd group, it was found that the anatomical distributions of the fluorescent PACT pathways along veins were the same as those of the enhanced PACT pathways by MRI (Figs. 1B1, 1C1, 1D1). When the thoracic cavity was opened, it was clearly visualized that the coronary, anterior and posterior grooves of the heart were stained by the fluorescein from the PACT pathways. By histological analysis and CLSM, the intrinsic structures of the PACT pathways were the outermost fibrous connective tissues covering the vessels from the extremities into the heart, within which were the longitudinally assembled and cross-linked fibers (Figs. 1B2, 1C2, 1D2, 1B3, 1C3, 1D3, 1E1-3). The sampled samples of the PACT pathways were examined by TEM and SEM and confirmed to be the cross-linked fibers at a micrometer scale (Figs. 1F1, 1F2). These findings coincided with our previous findings [1-3].

*ISF flow through the PACT pathways differed from blood flow through intraluminal cavity*

To illustrate the differences between ISF flow in a PACT pathway and the intraluminal blood flow, two types of angiography were used. Firstly, a segment of GSV was exposed and an incision in the PACT pathway was made in the 3rd group. When 20-30 µl FluoNa solution entered and mixed with blood in a branch of saphenous veins, the intravascular blood was partially stained by the fluorescein in which a spiraled and rotated flow pattern of blood flow was observed inside the lumen (Video 2). Whereas the incision of the PACT pathway had not been stained yet.

Secondly, after a fluorescent longitudinal channel in the PACT pathway was clearly displayed in the 9th group (Figs. 2A, 2B), 2 mL FluoNa solution was injected via ear vein in 2 subjects. The intraluminal FuoNa had visualized the blood in femoral artery (Fig. 2C) and GSV (Fig. 2D).

Thus, the demonstrated ISF flow through a PACT pathway was significantly distinguished from arterial and venous blood flow. Moreover, blood flow through vascular conduit is considered as laminar flow at low velocity, laminar-turbulent transitional flow, and turbulent flow at high velocity. Interestingly, our data suggested that the outermost layer of blood flow column was spiraling and rotating inside vascular conduit like a fired bullet through gun-barrel with periodic internal spiral ridge. One flow circle that blood spiraled and rotated inside the vascular vessel was found to be 1.3-1.9 sec. The flow length of one circle of the blood spiraling was calculated to be 5-8.4 mm given a speed of 6-7 cm/sec and a diameter of 2mm of the femoral venous blood flow [9] (Video 2). We predicted that there would be a periodic internal spiral ridge-like structure in tunica intima of vascular vessels and need further studies to verify.

*The calculated ISF flow speed through the PACT pathways by fluorescence stereomicroscopy*

To estimate the velocity of ISF flow through a PACT pathway, the FluoNa flow was recorded by real-time fluorescence stereomicroscopy in the 2$^{nd}$ group. During the initial observation, it was too difficult to catch the front end of the FluoNa flow along the vessels by stereomicroscopy. Fortunately, many fluorescent spots could be found to emerge and move along the PACT pathways in a 50-60 min observation when the gross fluorescence intensity began to unsymmetrically decrease (Video 3). After 60-70 min of the observation, it was found that a few large clumps of fluorescein were continuously flushed away along the vessels as well (Video 4). The calculated speed of FluoNa spots was 3.6-15.6 mm/sec and the sizes around 10-20 μm, which might represent fluid flow along adventitial fibers (Figs. 3A1-2). The calculated speed of FluoNa clumps was 0.1-7.6 mm/sec and the sizes around 50-400 μm, which might represent a late stage of fluid flow between tunica adventitia and its covering fascia (Figs. 3B1-2). At the edge of the arch-shaped incision of covering fascia, it was clearly observed that FluoNa clumps flowed down between adventitia and its covering fascia (Video 5, 6).

*A multilayer ISF flow in a PACT pathway*

To observe the whereabouts of ISF flow along the vessel, an incision in a PACT pathway of the SSV was made in the 3$^{rd}$ group. The cutting plane of the arch-shaped incision by scissor included layers of covering fascia and partial tunica adventitia (Figs. 4A1, 4A2). When 0.2 mL FluoNa solution was injected hypodermically into ankle dermis, the deeper layer of the exposed tunica adventitia, the cutting plane of the SSV were stained step-by-step (Figs. 4B1, 4B2, 4C1, 4C2). Then, the fluorescent ISF emerged out of the arch-shaped incision continuously (Figs. 4D1, 4D2, Video 5). The successive staining processes of the incision plane clearly showed that ISF flowed continuously through tunica adventitia firstly and into its surrounding connective tissues finally (Figs. 4E1, 4E2, Video 6). When more fluid was injected into ankle dermis, fluid would be conveyed by the PACT pathways and diffuse into the cavity between adventitia and its covering fascia, forming a constant bulk-like flow along the PACT pathways toward the heart direction.

Therefore, the revealed successive processes of fluid flow through a PACT pathway indicated ISF flow through tunica adventitia firstly and perivascular connective tissues secondly. Within perivascular connective tissues, a bulk-like fluid flow could form between layers of perivascular fascia.

*Two types of driving forces for ISF flow in a PACT pathway: "Pull and Push"*

The driving force of fluid flow through a PACT pathway has not been understood yet. To observe the driving force for ISF flow, fluid movements along the PACT pathway were studied in a segment of a vein that was isolated by the sutures from the other surrounding tissues and exposed in air. The "push and pull" experiments disclosed at least two types of *in vivo* forces to drive ISF flow (Figs. 5A1-4, 5B1-4). Without these two forces, the fluorescein would simply diffuse around the original drop site (Figs. 5A2, 5B2). By either the "push" or the "pull", the fluorescein in the middle of the isolated segment would be driven centripetally and stain the downstream PACT pathway (Figs. 5A4, 5B4). The "push" force might be due to sufficient supplies of ISF from the upstream PACT pathway, which is generated from capillaries. The "pull" force might be due to the mechanical movements of the heart beatings and disclosed in our previous studies of the amputated lower legs and the cadavers as well. The structural framework of vasculature's adventitia is composed of the long-distance assembled fibers from the distal veins to the epicardium and the three grooves of the heart [3]. Via the topographically connected ITZs along these cross-linked fibers, ISF would be driven toward the heart.

In the 5 sacrificed rabbits of the $6^{th}$ group, the right and left ventricular apex were cut off and the ventricular chambers were opened to air. By the repeated cardiac compressions on the ventricular chambers, FluoNa was "pulled" centripetally and stained the tunica adventitial of the SSV, IVC and the anterior groove of the heart in the 3 rabbits with the intact PACT pathways. When the fibrous framework on the SSV was destroyed by collagenase type I, FluoNa could not be "pulled" by the cardiac compressions and stay in the injection site in the other 2 subjects. Therefore, the ISF transport depended on two factors at least: the mechanical heart beatings as a driving source and the long-distance PACT pathways with an intact fibrous framework. The transport pattern of ISF toward an active dynamic driving source has not been comprehensively studied and need more explorations.

*The estimated pore sizes in a PACT pathway for ISF flow*

In conventional physiology, fibrous matrix can be considered as porous medium. The compositions of the PACT pathways are mainly fibrous matrix. Therefore, the porous PACT pathways may have two types of pores: closed pores and connected pores accessible to flow [10, 11]. Revealed by CLSM, the long-distance PACT pathways were composed of abundant fluorescently stained fibers along the SSV, IVC and into the anterior groove. In the controlled subject in the 4$^{th}$ group, there were no signs of the fluorescently stained fibers in the PACT pathways along the veins into the heart. The stained fibers were the results of the fluorescent ISF flow through a space of an ITZ along the fibers. In a view of porous medium, the longitudinally connected ITZs through the PACT pathways were equivalent to the connected pores or pore throats for fluid flow.

To determine the pore size of the longitudinally connected ITZs in physiological conditions, we explored the permeabilities of two types of nanoparticles through the PACT pathways in the 7$^{th}$ and 8$^{th}$ group. Administrated into the upstream pathways, the 5, 10, 20 nm gold nanoparticles and the 30 nm PS microspheres were found in the downstream PACT pathways by TEM (Fig. 1G) or the real-time fluorescence stereomicroscopy. The 40, 60 nm gold nanoparticles and 100 nm PS microsphere were not detected in the meanwhile. Thus, the *in vivo* pore size of the connected ITZs along the fiber might be under 100 nm and accessible for a nanoparticle with 10-20 nm approximately, which coincided with our previous findings in human cadavers [3].

Given the understandings of interfacial science, fluid transport through such a nanoscale interfacial zone with diameters ranging from 10 nm to 100 nm would be enhanced by the interfacial effects/interactions and significantly different from blood or lymph transport through the macroscale vascular conduits by pressure gradients as well as diffusive transport [12, 13]. In addition, the PACTs pathways were composed of abundant collagenous fibers, within which there were a few elastic fibers (combined) (Fig. 1E3). Typically, the collagenous molecules were hydrophilic and rich in the amino acid sequences of glycine-proline-X and glycine-X-hydroxyproline. By comparison, the hydrophobic elastic fibers were rich in glycine and proline. Thus, the surfaces on the fibril bundles must contain hydrophilicity and hydrophobicity to some extent [14]. Here, we postulated that fluid transport through the PACT pathways would be impacted by a surfactant due to the potential hydrophobic and hydrophilic characters of the connected ITZs in a PACT pathway [15-17].

*Two parts in a PACT pathway: a longitudinal tunica channel part and an absorptive part*

In the 2[nd] group, the dynamic transport processes of ISF flow along blood vessels indicated there were two parts in a PACT pathway: a longitudinal tunica channel along the vascular vessel and an absorptive part around tunica adventitia. In the 9[th] group, the two parts of a PACT pathway were visualized respectively by the FluoNa SDS solution (Figs. 6B1-3). During the first 25-30 min observation, the imaging mixture of FluoNa SDS solution stained 2-3 tunica channels along the GSV and 1 tunica channel along SSV. These longitudinal tunica channels had a clear boundary and the absorptive part of the PACT pathways had not stained yet. After 30-40 min around, the absorptive part of GSV or SSV was gradually stained by the fluorescein from the tunica channel. At 60 min after the administration, the PACT pathways surrounding the vessels were totally stained like the findings in the 2[nd] group (Figs. 6A1-3). By CLSM, the PACT pathways in the 9[th] group were full of the fluorescently stained fibers as well.

The same results of the 9[th] group were found with a nonionic surfactant, Trixon-100 in the 10[th] group. For the cationic surfactant in the 11[th] group, the longitudinal tunica channel along the PACT pathways was visualized by CTAB but with a fuzzy boundary (Figs. 6C1-3). When the tunica channel was visualized, the absorptive part began to be stained gradually but unevenly. At 60 min after the administration, the PACT pathways surrounding the GSV or SSV showed a patchy distribution of fluorescein (Figs. 6C2, 6C3). By CLSM, the PACT pathways in the 9[th]-11[th] group were full of the fluorescently stained fibers as well.

These revealed macroscopic behaviors of adventitial ISF flow further illustrated there were two parts in a PACT pathway: a longitudinal tunica channel part and an absorptive part, like a fractured-porous rock that contains highly ordered polygonal pores connected to slit-shaped spaces or pore-throats [10, 11, 18]. The tunica channel part was similar with a pore throat accessible for fluid transport through porous medium. The absorptive part was equivalent to the closed pores that attract free water into porous medium. The tunica channel part might be related with the stained ITZs along the longitudinally assembled fibers because neither the fibers nor the tissue gel can flow [8]. The absorptive part might be associated with the gel-like ground substances of adventitial matrix, which was a highly absorbent biological gel [6].

According to the properties of surfactants, we made a speculation [14-18]. Surfactants are compounds that lower surface/interfacial tension between two phases. When mixed with a surfactant, the interfacial tension of fluid in contacting the gel was lowered and the absorption of the fluorescent

ISF into the gel would be suspended. In the 9$^{th}$, 10$^{th}$ groups with the anionic/nonionic surfactants, the fluorescent ISF into the absorptive part had been significantly postponed (Figs. 6B1-3). However, in the 11$^{th}$ group with cationic surfactant, the fluorescent ISF into the absorptive part had not been completely stopped (Figs. 6C1-3). The reason may be due to the positive charged cationic surfactants were attracted by the negative charged gel of adventitial matrix and weaken their abilities to block the fluorescent ISF into the absorptive part [6, 19]. By contrast, the anionic/nonionic surfactants are negative or neutral charged and repulsive to the adventitial matrix.

*ISF flow through a longitudinal tunica channel in a PACT pathway*

To explore ISF flow through a longitudinal tunica channel in a PACT pathway, two types of experiments were performed. Firstly, the peripheral ISF via the PACT pathways into the pericardial cavity was examined between the FluoNa water solution and the FluoNa SDS solution. In the 12$^{th}$ group with the hypodermic injection of 3-4 mL FluoNa water solution, extra pericardial fluid was found by echocardiography in approximately 40-60 minutes. By the hypodermic injection of 3-4 mL FluoNa SDS solution (contained 5% SDS), there was no sign of pericardial effusion found in the 13$^{th}$ group. The amount of ISF flow through the tunica channel part of the PACT pathways was decreased by the surfactant. Secondly, the ISF flow through a PACT pathway was tested by 5%, 10% and 20% SDS solution respectively in the 14$^{th}$ group. Accompanying with the increased surfactant concentration, ISF transport through a PACT pathway was finally suspended in 20% SDS solution. Although studies on the detailed mechanisms are needed, the phenomenal findings strongly suggested the involvements of interfacial interactions in ISF flow through the tunica channel part and absorptive part of a PACT pathway [11].

*A working hypothesis for ISF dynamic transport*

Based on the current data and our previous findings on the periodic to-and-fro "gel pump", the dynamotaxis, a longitudinally assembled and topologically connected ITZs through fibrous connective tissues [3, 7, 8], a working hypothesis for ISF dynamic transport mechanism was proposed: 1, Connective tissues provide the gel-like ground substances to create an interfacial zone paving on a solid surface everywhere in the body, such as a fiber, a cell, a bundle of fibers, a group of cells, layers of fascia and other solid surfaces, consisting of a multiscale, multilayer and multiform space ranging from nano-micron-sized thickness to meter-sized length in biological system. 2, ISF flow through a network of interstitial connective tissues depends on two essential factors: a "gel pump" and a long-distance interconnected interfacial zone, the two parts of which are topologically connected. 3, The "gel pump" is the gel-like fibrous matrices of connective tissues in a driving source/center, working as a device that drives fluids by mechanical actions. When a tissue or an organ is compressed by the to-and-fro mechanical movements, the gel in the driving source/center becomes a "drive pump" by attracting and releasing fluid into and out of the gel. 4, When fluid comes into the long-distance interconnected interfacial zone, fluid flow would be in response to a dynamic driving source/center and usually "pushed" or "pulled" toward the gel pump. 5, The interfacial interactions of fluid with the solid surfaces and the contacting gel might be a dominated driving force to push or pull fluid flow along a long-distance and topologically connected ITZ as well as absorb fluid into the gel [20]. 6, Theoretically, the absorptive processes of fluid into the gel via the ITZs would be more efficient than the diffusive processes of fluid into the gel without the help of an ITZ. When the gel is compressed, fluid would be squeezed out of the gel into the surrounding environments rather than into the ITZs. When the gel is decompressed, fluid in the ITZs would be absorbed easily into the gel and resupplied by the fluid from a neighboring ITZ. Consequently, fluid in the distal neighboring ITZs would enter the proximal neighboring ITZs by interfacial forces. A continuous interfacial fluid flow might be established by the continuous fluid resupplying processes throughout the long-distance and topologically connected ITZs. When the gel is compressed repeatedly, fluid in a far distance would be conveyed into the "gel pump" and extra fluid in the gel would be "pumped" into the sac of the driving source/center as well as the cavity covering the long-distance ITZs pathways. 7, A biotic interfacial fluid flow system might provide a physical field between matter phases to transport the physical/chemical signals/information/energy and regulate the physiological functions throughout the whole body by mechanical driving forces and other synergistical dynamic powers.

In the human cadaver experiments, the peripheral ISF in the thumb had been driven into the superficial tissues on the heart that was repeatedly compressed by an automatic cardiac compressor [3]. In alive rabbits, the peripheral ISF was conveyed by a PACT pathway into the tissues on the heart and entered the pericardial cavity, causing pericardial effusion [1]. In the current surfactant studies, we found that the long-distance transport processes of ISF in a PACT pathway were suspended by 20% SDS solution. Presumably, the interfacial interactions of fluid with the solid surfaces and the contacting gel along a long-distance and topologically connected ITZs might be damaged by the increased amount of the surfactants. Of course, the exact dynamic mechanisms of interfacial interactions for ISF flow need further studies.

*ISF in the PACT pathways converged into capillaries and venous vessels nearby*

As to the relationship with blood circulation, it was easily found that the fluorescent ISF in a PACT pathway has been retaken by the capillaries into venous vessels nearby (Figs. 7A1-2, 7B, Video 6). The results coincided with the understanding of ISF exchange with capillaries that was described by Starling equation [21-23]. Thus, the continuous ISF flow in a PACT pathway can exchange constantly with blood circulation via the capillaries along the PACT pathways as well as the capillaries of coronary circulation.

*The respiratory movement of lung was another potential driving center for ISF flow*

In the 4$^{th}$ group, it was displayed by CLSM that the stained covering fascia on the lung and the PACT pathways of the pulmonary vein were composed of longitudinally assembled and cross-linked fibers as well (Figs. 8). The respiratory movements seemed to be another driving center to "pull" ISF from the PACT pathways.

In summary, the current studies further revealed one of a long-distance ISF flow network that is a continuous fluid flow along the vasculature in an animal, including ISF flow through tunica adventitia and fluid flow through an interfacial zone between adventitia and its covering fascia. Our data revealed a successive process of ISF flow through a PACT pathway, which was the adventitia firstly and perivascular connective tissues secondly. When sufficient fluid come into the PACT pathways, extra fluid would diffuse into the cavity between adventitia and its covering fascia and flow constantly. Because neither fibers nor gel substances can flow, an interfacial zone between a solid surface and a gel-like substance was suggested as a space for ISF flow. The space size of interfacial zone is not few layers of molecules but nano-sized or micron-sized in transverse section and macroscopic scale in longitudinal plane. Along tunica adventitia of the vasculature, an interfacial zone along adventitial fibers or between adventitia and its covering fascia was longitudinally assembled and topologically connected to constitute a macroscopic space throughout the PACT pathways and into the surfaces of the heart and lung. When fluid comes into the interfacial zones, ISF transport in a PACT pathway would be a continuous flow or even a bulk-like flow by a dynamic driving source/center. We named this flow pattern as a biotic "interstitial/interfacial fluid flow". Without extra active dynamic driving power, ISF diffuses locally. Our presented data strongly suggested the interfacial interactions play an essential role in both parts of a "slit-shaped" PACT pathway for ISF flow. By the ingenious usages of interfacial tension, an active dynamic source/center could regulate ISF circulation locally between cells and vasculature as well as systemically around the whole body. The demonstrated interstitial/interfacial fluid flow might be another principal fluid dynamic pattern besides convective/vascular and diffusive transport in biological system. In a view of interfacial science, we predicted there would be a multiscale, multilayer, and multiform interfacial fluid flow in vascular walls as well as other various tissues and organs containing an interfacial zone between two phases of biomaterials. As to fluid flow through an arterial PACT pathway, more studies are needed. The components conveying by an interstitial/interfacial fluid flow system need to be verified, including fluid, bio-signal molecules, nutrients, solutes, oxygen, metabolic products, electrical ITZ potential, electromagnetic signals, drugs, etc. To comprehensively understand the physiological and pathophysiological functions of a PACT pathway along vasculature and other ISF circulation is an extremely intriguing subject for future studies in life science.

Acknowledgments

This study was supported by National Natural Science Foundation of China (No. 81141118), Beijing Hospital Clinical Research 121 Project (121-2016002) and National Basic Research Program of China (Nos. 2015CB554507). We thank Ms. SiuTuen Lucy Chan Lau, Mr. WaiChun Tin and Weiwu Hu for their financial support and Professor Saicang Lobsang Hwardan Chegynima, Jamyang Danchour, Angwenzhabar, Peisu Xia, and Hua Li for their help with data analysis, Wentai Lee for his assistance with the English language corrections of this manuscript.

Author Contributions

HyL conceived the original ideas, conceptions and designed the experiments. HyL, LX, XlC, YL performed the experiments. HyL, QH, FsJ, HL analyzed the data. HyL contributed reagents/materials. HyL, YjY analysed the interfacial transport pattern. HyL, BL prepared pictures and videos. HyL wrote the paper.

Disclosure

The authors declare no competing financial interests and have no conflicts to disclose.

**Figures and figure legends**

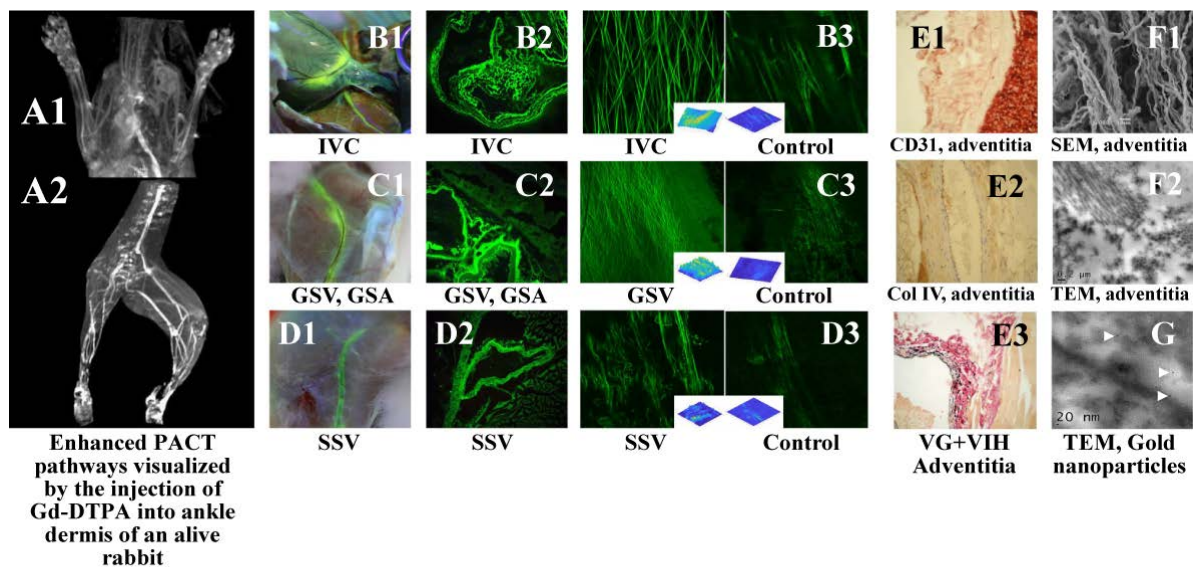

**Figure 1. Illustrations of the PACT pathways. A1-2** showed the panorama of the enhanced long-distance PACT pathways originating from ankle dermis in 1$^{st}$ group by MRI. **B1** showed a segment of IVC and coronary groove were stained by FluoNa from ankle dermis in 2$^{nd}$ group. **B2** was cross-sectional view of a sample sampled from **B1** by frozen microscopy and showed the vessel wall and a venous valve were stained. **B3** showed abundant fluorescently stained and cross-linked fiber in venous adventitia by CLSM, which were running longitudinally toward the long axis of the vein and sampled from **B1**. **C1** showed a segment of great saphenous vein (GSV) and artery (GSA) were stained by FluoNa from ankle dermis in 2$^{nd}$ group. **C2** was cross-sectional view of **C1** and showed

the venous and arterial adventitia and their surrounding tissues were stained. **C3** showed fluorescently stained fibers in venous adventitia of **C1** by CLSM. **D1** showed a segment of small saphenous vein (SSV) was stained by FluoNa from ankle dermis. **D2** was cross-sectional view of **D1** and showed the venous adventitia and its surrounding tissues were stained. **D3** showed fluorescently stained fibers in venous adventitia of **D1** by CLSM. **E1** (×40, CD31), **E2** (×20, collagen IV) showed there were no appearances of numerous ring-like groupings or linear structures within the venous adventitia and its surrounding loose connective tissues by immunohistochemical staining. **F1** (SEM), **F2** (TEM) showed the venous adventitia contained abundant fibers. **G** showed gold nanoparticles of 5, 10 and 20 nm diameters that were located among the fibril bundles (white arrows) of the downstream PACT pathway in the 7$^{th}$ group.

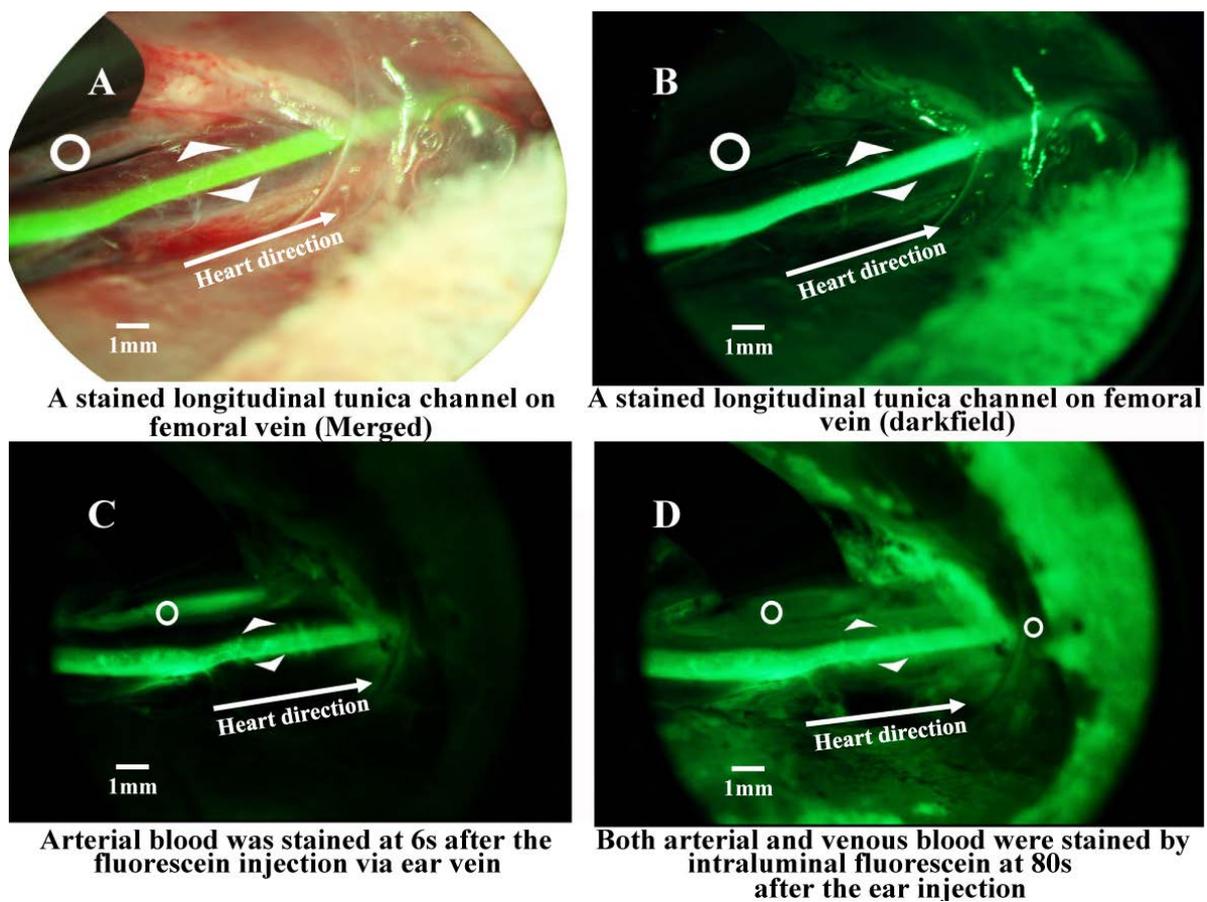

**Figure 2. Illustrations of the differences between FluoNa flow through the PACT pathway and intravascular cavity**. **A** showed a longitudinal tunica channel in the PACT pathway of GSV that

was stained by FluoNa from ankle dermis in the 9[th] group (Merged of darkfield and brightfield). **B** was the darkfield of **A**. **C** showed the femoral arterial blood (pointed by white circle) was stained at 6 second after intravascular injection of FluoNa into ear vein. **D** showed both the femoral arterial blood and venous blood were stained by intravascular FluoNa while the longitudinal tunica channel in the PACT pathway was clearly stained in venous adventitia of femoral vein. Thus, ISF flow through a PACT pathway was significantly distinguished from arterial and venous blood flow.

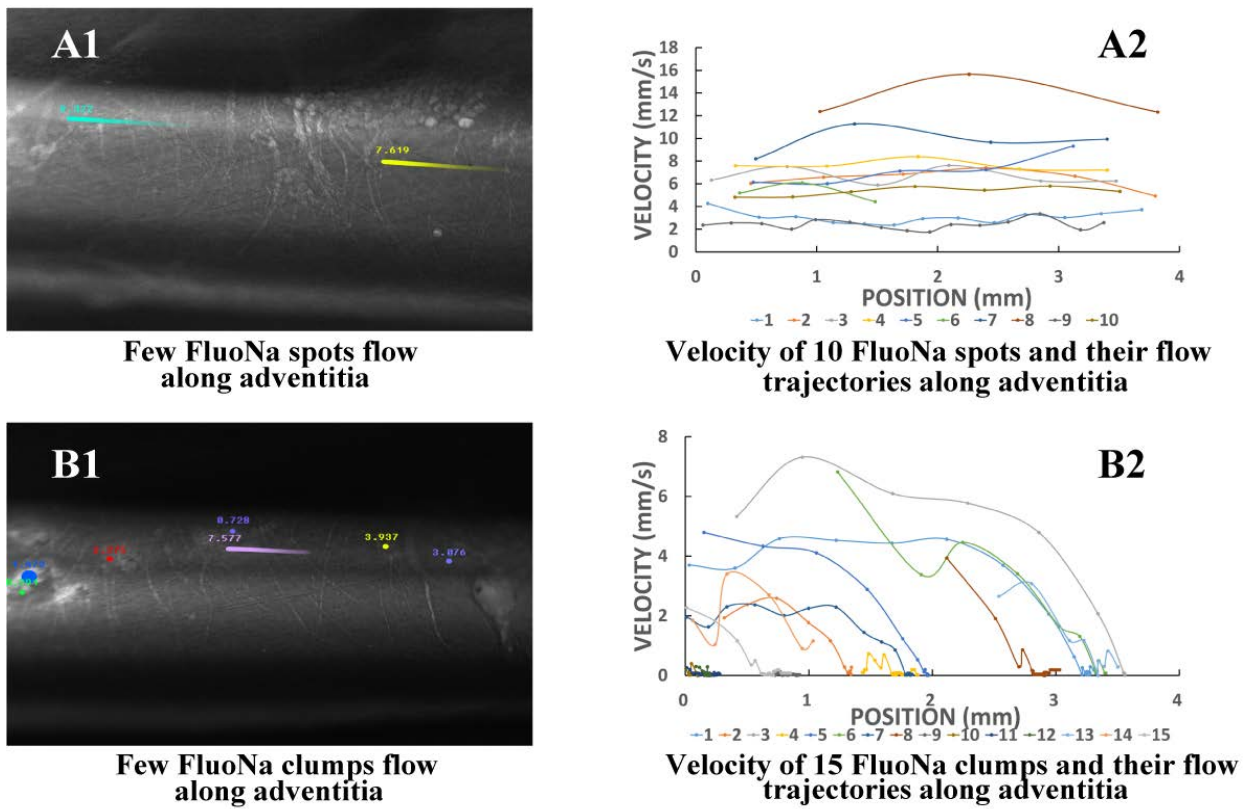

**Figure 3. Velocity of few FluoNa spots or clumps flow along adventitia. A1** showed 2 FluoNa spots flow trajectories along a segment of the PACT pathway in a 50-60 min observation (Video 3). After 60-70 min of the observation, it was found that a few large FluoNa clumps continuously flushed away along the vessels (**B1**). **A2** showed the velocity of a total of 10 FluoNa spots and their flow trajectories along the adventitia. The calculated speed of FluoNa spots was 3.6-15.6 mm/sec and the sizes around 10-20 μm, which might represent fluid flow along fibers. **B1** illustrated 7 FluoNa clumps flow trajectories along adventitia (Video 4). **B2** showed the velocity of a total of 15

FluoNa clumps and their flow trajectories along the adventitia. The calculated speed of FluoNa clumps was 0.1-7.6 mm/sec and the sizes around 50-400 μm, which might represent fluid flow between tunica adventitia and its surrounding fascia (Video 5, 6).

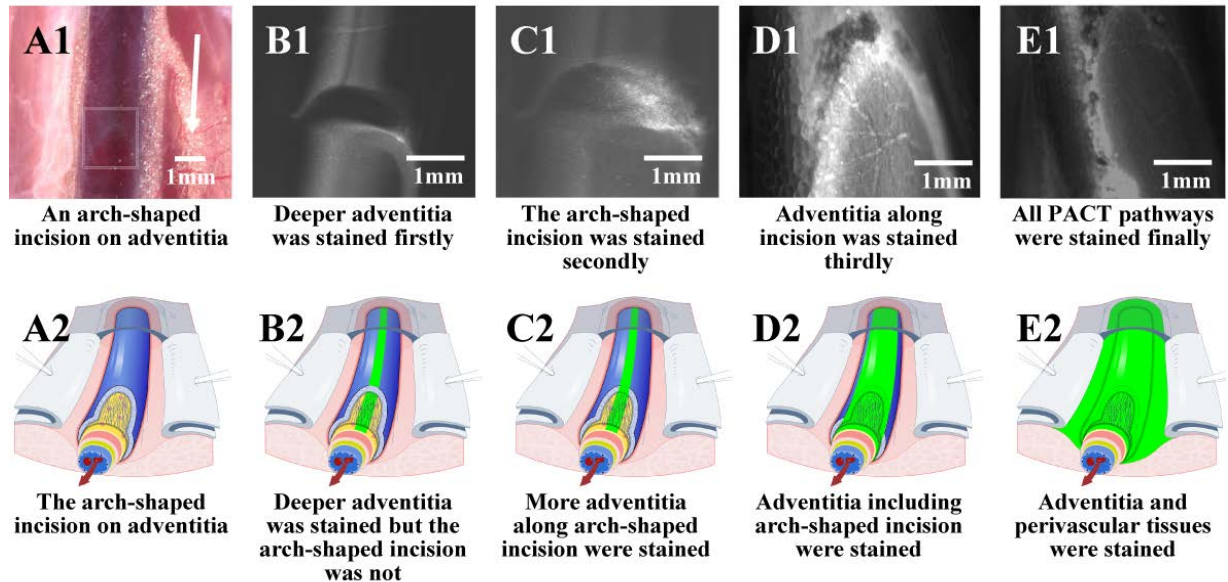

**Figure 4. Illustrations of successive processes of ISF flow through the PACT pathway**. **A1** showed an arch-shaped incision of tunica adventitia of a venous vessel (white box). **A2** was the diagram from **A1**. After the hypodermic injection into ankle dermis, FluoNa flowed along the PACT pathway on the vein and has firstly stained the deeper layer of adventitia but the arch-shaped incision was not stained (**B1**, **B2**). Secondly, the arch-shaped incision was gradually stained by FluoNa from ankle dermis (**C1**, **C2**). Thirdly, both deeper and superficial layers of adventitia around the incision were stained (**D1**, **D2**). When sufficient fluid come into the PACT pathways, the entire tunica adventitia and the surrounding tissues around the vessel were stained (**E1**, **E2**). In the meanwhile, extra fluid would diffuse into the interfacial zone between adventitia and its covering fascia and flow down along the edge of the incision (**Video 5, 6**). The demonstrated successive processes of FluoNa flow through the PACT pathway indicated that the long-distance adventitial ISF flow was through tunica adventitia firstly and perivascular connective tissues secondly.

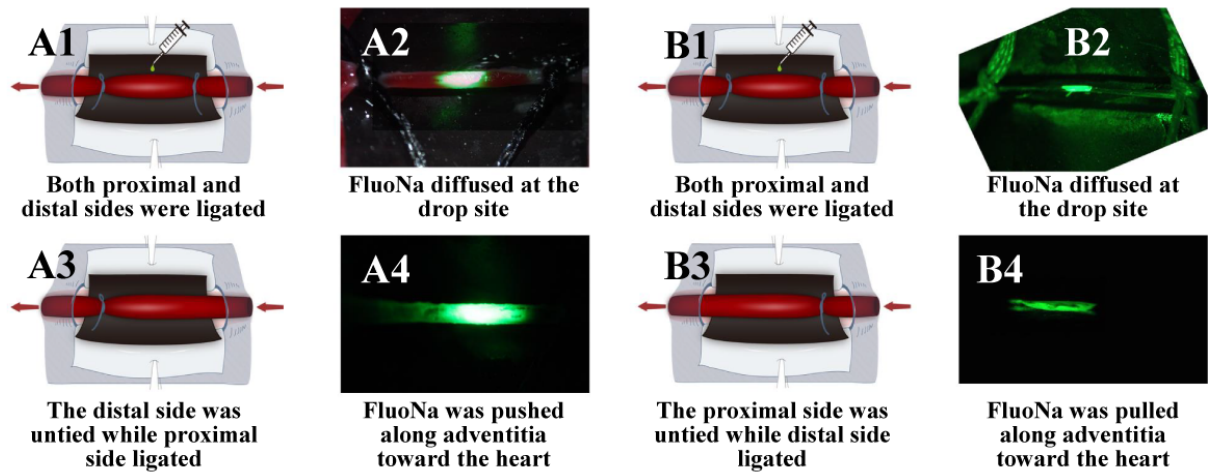

**Figure 5. Two types of ISF flow through a PACT pathway demonstrated by FluoNa surfactant solution**. At 10 min after the injection of 0.1mL FluoNa water solution into ankle dermis, the entire tunica adventitia around the vein were stained (**A3** merged by **A1** and **A2**). At 10 min after the injection of either 0.1 mL FluoNa SDS (anionic surfactant) solution or 0.1mL FluoNa Trixon-100 (nonionic surfactant) solution into ankle dermis, a longitudinal tunica channel was clearly visualized within the PACT pathway of venous vessel (**B3** merged by **B1** and **B2**). Usually, there were 2-3 tunica channels in the PACT pathway of GSV and 1 tunica channel on SSV. At 10 min after the injection of 0.1 mL FluoNa CTAB (cationic surfactant) solution into ankle dermis, a patchy distribution of FluoNa around the entire PACT pathway on GSV was displayed (**C3** merged by **C1** and **C2**). The revealed macroscopic behaviors of adventitial ISF flow suggested there were two parts in a PACT pathway: a longitudinal tunica channel part for a long-distance transport and an absorptive part for local diffusive transport.

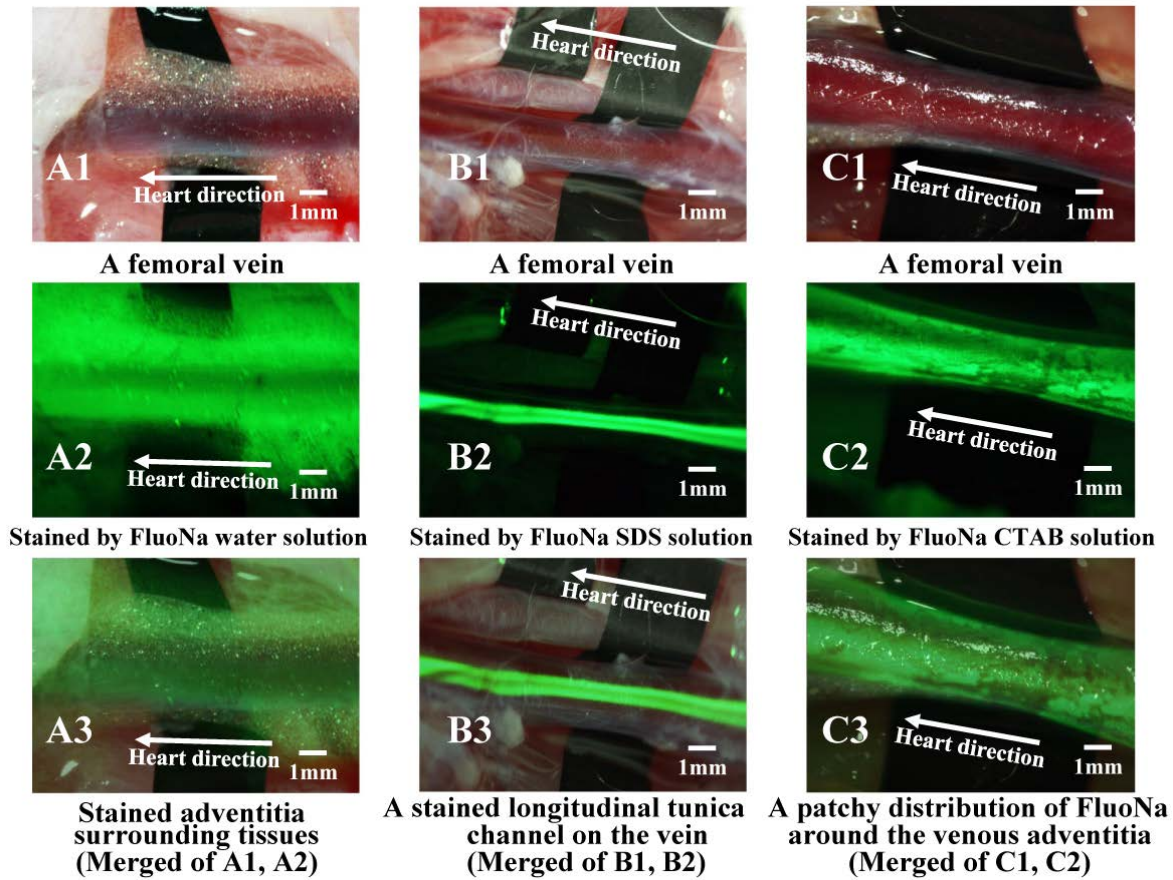

**Figure 6. Illustration of two types of driving forces to "push and pull" ISF flow in a PACT pathway. A1, B1** illustrated that a segment of venous vessel was ligated sutures. A2, B2 illustrated that FluoNa in the middle of the ligated segment diffused locally at the drop site when both the proximal and distal sides were ligated. **A3, A4** showed FluoNa was "pushed" by the force from the distal side when the distal side was untied. **B3, B4** showed FluoNa was "pulled" by the force from the proximal side when the proximal side was untied. Without these two forces, FluoNa would simply diffuse around the original drop site. By either the "push" or the "pull", the fluorescein in the middle of the isolated segment would be driven centripetally and stain the downstream PACT pathway.

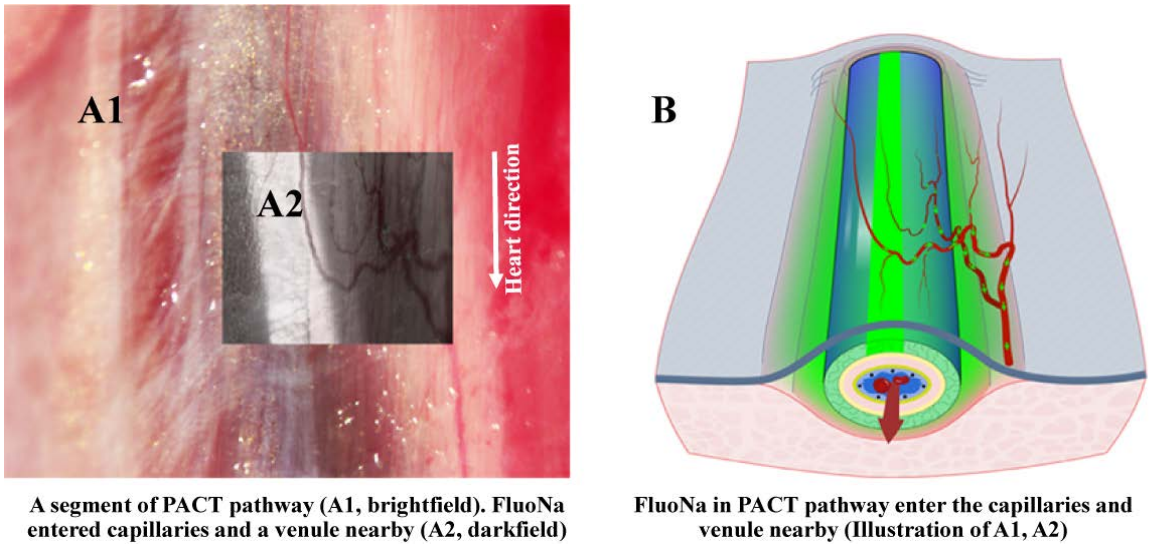

**Figure 7. Illustration of FluoNa in a PACT pathway on a vein entered the capillaries nearby.** **A1** showed a segment of venous vessel and capillaries. **A2** was darkfield of part of **A1** and showed FluoNa in the PACT pathway entered capillaries and a venule nearby (Video 6). **B** was the diagram of **A1** and **A2**. These findings coincided with Starling equation and indicated that an adventitial ISF flow would exchange constantly with blood circulation via the capillaries along the PACT pathways.

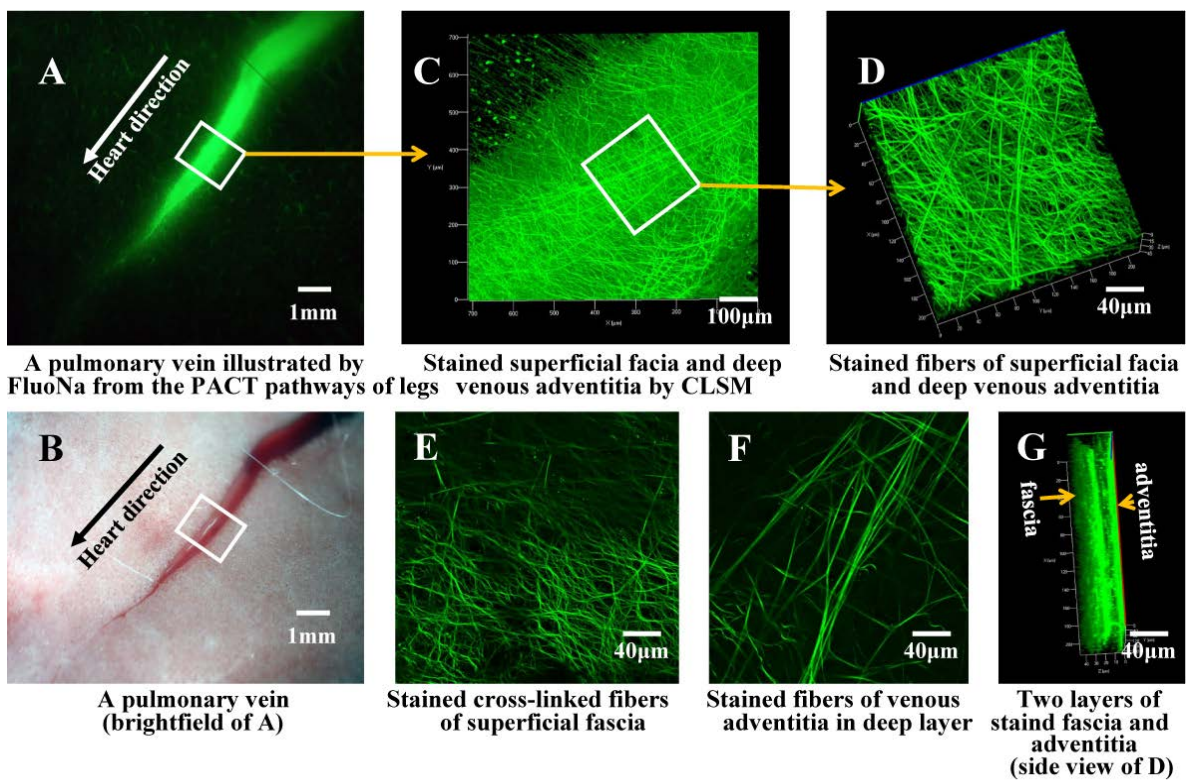

**Figure 8. Illustration of a segment of pulmonary vein found in the 4$^{th}$ group that was

**sacrificed at 15 min after the hypodermic injection**. **A** showed a pulmonary vein seemed to be stained by FluoNa from the ankle dermis. **B** was brightfield of **A**. **C** showed there were two types of fibers on the venous vessel found by CLSM. **D** was 3D view of enlarged **C**. **E** showed the cross-linked fibers in the superficial layer of **D** that were stained by FluoNa from ankle dermis. **F** showed the longitudinal fibers of adventitia in the deep layer of **D**. **G** was the side view of **D** and showed two layers of stained fibers that were located in the superficial fascia of the lung and the deep layer of venous adventitia.

**Video 1. Illustration of the long-distance PACT pathways that was scanned at 40 min after the hypodermic injection into ankle dermis by MRI**. Originating from ankle dermis, the paramagnetic tracer of Gd-DTPA seemed to have enhanced the venous vessels of the limbs, the abdominal and thoracic cavities and into some tissues of the heart. Limited by MRI technique, whether Gd-DTPA was inside the venous cavity or in the PACT pathways outside the vessels cannot be determined. Compared with the findings by fluorescence microscopy, the enhanced pathways by Gd-DTPA coincided with the fluorescently stained PACT pathways by FluoNa from ankle dermis.

**Video 2**. **Illustration of dynamic flow of partial stained blood in a vein by fluorescence stereomicroscopy**. After 20-30 μl FluoNa solution entered and mixed with the blood in a small branch of the upstream veins in the 3$^{rd}$ group, the intravascular blood was partially stained by FluoNa. Under the real-time observation, a spiraling and rotating flow pattern of blood flow was clearly visualized inside the vascular vessel. In the meanwhile, the arch-shaped incision of the PACT pathway on the vein has not been stained yet. One flow circle that blood spiraled and rotated inside the vascular vessel was found to be 1.3-1.9 sec. The flow length of one circle of the blood rotating was calculated to be 5-8.4 mm given a speed of 6-7 cm/sec and a diameter of 2 mm of the femoral venous blood flow. We predicted that there would be a periodic internal spiral ridge-like structure along tunica intima of vascular vessels and need further studies to verify.

**Video 3**. **Illustration of FluoNa spots movements along the PACT pathway dynamically recorded by fluorescence stereomicroscopy**. Around 50-60 min after the injection of FluoNa into ankle dermis, a few FluoNa spots were found to emerge and move along the PACT pathway. The calculated speed of FluoNa spots was 3.6-15.6 mm/sec and the sizes around 10-20 μm, which might represent fluid flow along adventitial fibers (Figs. 3A1-2).

**Video 4**. **Illustration of FluoNa clumps movements along the PACT pathway dynamically recorded by fluorescence stereomicroscopy**. Around 60-70 min after the injection of FluoNa into ankle dermis, a few FluoNa clumps were continuously flushed away along the vessel wall. The calculated speed of FluoNa clumps was 0.1-7.6 mm/sec and the sizes around 50-400 μm, which might represent a late stage of fluid flow between tunica adventitia and its surrounding fascia (Figs. 3B1-2).

**Video 5**. **Illustration of a continuous ISF flow in an arch-shaped incision of the PACT pathway by real-time fluorescence stereomicroscopy**. Dynamically observed after the injection of FluoNa into ankle dermis, it was clearly visualized that ISF flowed along the PACT pathway and emerged

outside the arch-shaped incision. The dynamic processes of continuous and successive adventitial ISF flow were showed in Figure 3.

Video 6. **Illustration of a continuous fluid flow along the edge of an arch-shaped incision of the PACT pathway by real-time fluorescence stereomicroscopy**. After ISF flow was clearly visualized in the arch-shaped incision (Video 5), extra fluid was found to flow down along the edge of the arch-shaped incision and the exposed outermost layer of adventitia. These dynamic processes indicated the space for extra fluid flowing was an interfacial zone between the adventitia and its covering fascia.

**Video 7. Illustration of adventitial ISF flow converged into capillaries and a venule nearby.** During the adventitial ISF transport courses along the PACT pathway, FluoNa was found to converge into the capillaries and a venule nearby. Thus, the continuous ISF flow in a PACT pathway can exchange constantly with blood circulation via the capillaries along the PACT pathways.